\documentclass{iau} 
\usepackage{graphicx}
\usepackage[table]{xcolor}
\title[JD 11.~~pulsars and transients with GMRT] 
{Search for pulsars and transients with the GMRT}

\author[Bhaswati Bhattacharyya]  
{Bhaswati  Bhattacharyya$^1$}

\affiliation{$^1$NCRA-TIFR, \\ Pune University Campus,
Ganeshkhind, Pune - 411007 \\ email: {\tt bhaswati@ncra.tifr.res.in} \\[\affilskip]
}

\pubyear{2017}
\volume{337}  
\jname{Pulsar Astrophysics -­ The Next 50 Years}
\editors{P. Weltevrede, B.B.P. Perera, L. Levin Preston \& S. Sanidas, eds.}

\begin{document}

\maketitle

\begin{abstract}
This paper details on the discovery of 21 pulsars using the Giant Metrewave Radio
Telescope (GMRT) from targeted (Fermi directed search) and blind surveys (GMRT High Resolution Southern Sky - GHRSS)
and results from the follow up studies. We discovered seven millisecond pulsars (MSPs) in the Fermi
directed searches, which are the first Galactic MSPs discovered with the GMRT.
We have discovered 13 pulsars (including a MSP and two mildly recycled pulsars) with the GHRSS survey,
which is an off-Galactic-plane survey at 322 MHz with complementary target sky (declination range $-$40 deg
to $-$54 deg) to other ongoing low-frequency surveys by GBT and LOFAR.
The simultaneous time-domain and imaging study for localising
pulsars and transients and efficient candidate investigation with machine learning are some of the
features of the GHRSS survey, which are also finding application in the SKA design methodology.

\end{abstract}

\firstsection 
\section{Introduction}
Pulsars still have tremendous untapped potential to probe the behaviour of matter, energy, space and
time under extra-ordinarily diverse conditions. 
Investigation of single pulse behaviour of normal pulsars having spin period $>$ 30 ms
can exhibit interesting individual properties, like glitches, profile state changes,
nulling and intermittency. 
Stability, compactness second only to black holes, and their presence in binary systems, makes 
millisecond pulsars (MSPs) ideal laboratories to test the physics of gravity and
as detectors for long-wavelength gravitational waves (GWs). Millisecond pulsars (MSPs) are still a 
small population compared to the normal pulsars and much diversity in their intrinsic characteristics 
as well as evolutionary history are yet to be explored.
\begin{figure}[b]
\vspace*{-1.0 cm}
\begin{center}
 \includegraphics[angle=-90,width=5.5in]{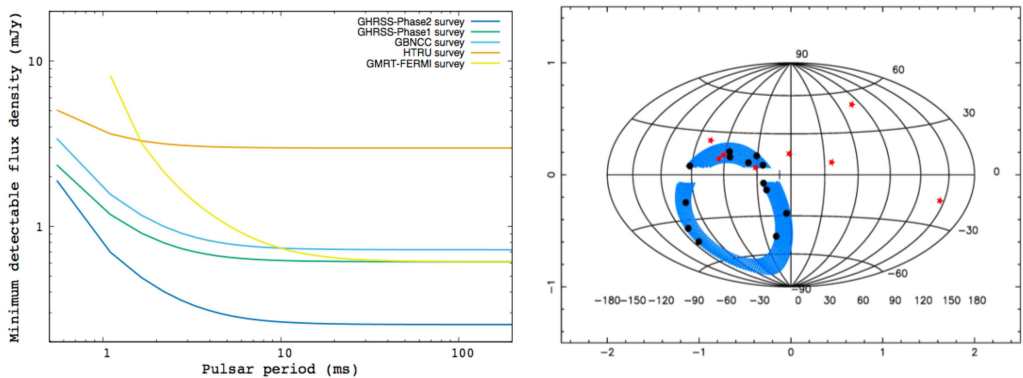}
 \vspace*{-2.0 cm}
 \caption{Left panel: Estimated sensitivity of the Fermi-directed search with the GMRT, GHRSS survey Phase 1, GBNCC and the HTRU survey. Right panel: Pulsars discovered by GMRT between 2012$-$2017 from targeted and blind surveys. Fermi-directed discoveries $-$ stars, GHRSS discoveries $-$ solid circles, sky coverage of GHRSS survey $-$ shaded region.}
   \label{fig1}
\end{center}
\end{figure}
Even though pulsars are frequently getting discovered with ongoing surveys at major telescopes over the 
world, presently known population is only 1\% of prediction. When blindly searched in the 
radio band, the MSPs are difficult targets due their intrinsic faint nature, and their detection 
generally requires long exposures even with large radio telescopes. So targeted searches with apriori 
knowledge of possible position is a effective way to search for MSPs (\cite{ferrara17}). We have carried 
out Fermi-directed search for MSPs with the GMRT, as a part of Fermi pulsar search consortium.
 
However, since one does not have apriori information about position of majority of the pulsars sensitive 
blind surveys are being designed to detect those. We are conducting the GMRT High Resolution Southern 
Sky (GHRSS) survey for pulsars and transients with the GMRT, targeting a portion of sky complementary to other 
ongoing low-frequency surveys.

In addition to the regular emission from pulsars, highly dispersed fast radio bursts 
(FRBs) of extragalactic origin are also detected in time domain surveys.  
 Moreover, low-frequency detection of FRBs is important considering the fact that no FRB is detected below 800 MHz. 
\section{Time domain study with GMRT}
The GMRT is a multi-element aperture synthesis telescope consisting of 30 antennas, each of 45 m diameter, spread over a 
region of 25 km diameter and operating at 5 different wave bands from 150 MHz to 1450 MHz. The GMRT 
Software Backend, built using COTS components, is a fully real-time backend that supports a FX correlator and a beamformer 
for an array of 32 dual polarized signals, Nyquist sampled at 33 or 66 MHz (\cite{Roy10}). Time domain study with the GMRT is 
aided by it's sensitivity and simultaneous imaging capability. 
\subsection{Sensitivity}
As pulsars have steep radio spectra the GMRT is ideal for searches away from the Galactic plane. 
Right panel of Figure \ref{fig1} plots the estimated sensitivity calculated with the radiometer equation for the Fermi directed survey at GMRT, 
GHRSS survey Phase-1 (with 32 MHz bandwidth), the GBNCC at GBT and the HTRU at Parkes. The sensitivity
estimations are at 322 MHz for 5-$\sigma$ detection with 10\% duty cycle in cold sky. The flux densities 
are scaled using $-$1.4 spectral index to derive the corresponding 322 MHz values. For Fermi-directed survey 
with 512 frequency channel over 32 MHz observing band (frequency resolution $\sim$0.062 MHz, time resolution $\sim$61 microsecond) at 
322 MHz we calculate 5-$\sigma$ detection sensitivity of 0.9 mJy for 30 min pointing. Sensitivity 
is improved by $\sim$ 2.5 times with enhanced time and frequency resolution modes (0.015 MHz, 61 microsecond) developed for the 
GHRSS survey. Calculated theoretical sensitivity for 10 mins pointing duration for the GHRSS at 322 MHz is 0.5 mJy for 32 MHz
observing band-width. 
\subsection{Image Domain Localisation}
Discovery position of pulsars and transients are subject to large uncertainty of the observing beam (e.g. $\pm$ 40$'$ 
for GMRT$-$322, $\pm$ 18$'$ for GBT$-$350, $\pm$ 7$'$ for Parkes$-$Lband) hinder sensitive follow up studies of these newly 
discovered objects using coherent beams of array telescopes, or at higher frequencies using single dishes. 
We can localise the newly discovered pulsars and transients in the image plane with the GMRT interferometric array with an 
accuracy of better than $\pm$ 10$''$ (half of the typical synthesized beam used in the image made at 322 MHz) using gated 
imaging of pulsars (\cite{Roy13}).
\section{Fermi directed search for MSPs with the GMRT}
Using the GMRT we performed deep observations to
search for radio pulsations in the directions of unidentified Fermi Large Area
Telescope (LAT) $\gamma$-ray sources. With the aid of reduced quantization noise and
high time resolution survey mode, the GMRT software backend (GSB) enhanced the
search sensitivity for nearby MSPs by $\sim$ 80\% than what was previously possible. 
The capability of the GMRT in finding MSPs, even in tight binaries, is demonstrated by the discovery of seven
MSPs in Fermi directed searches in mid and high Galactic latitudes (\cite{Bhattacharyya_etal2013}). 
These include the first Galactic field MSPs discovered at the GMRT.
Two of the discoveries are serendipitous in-beam MSPs at a large offset from
the search centers, even outside the half power beam width. Some of the objects that
the GMRT has discovered, are already pushing the boundaries of the known parameter
space of MSPs; for example, J1544$+$4947 $-$ eclipsing Black-Widow MSP (\cite{Bhattacharyya_etal2013}) 
and J1536$-$4948 $-$ a MSP having very wide pulse profile (possibly widest among MSPs). We have also discovered 
J1227$-$4853 $-$ a transitional MSP caught in the act of transition from a Low-mass X-ray Binary to radio MSP.
Parameters of these MSPs are listed in Table \ref{tab1}. 
While folding data with radio timing model, LAT pulsation is detected from 4 of the MSPs discovered by the GMRT.
Since two of the MSPs are unassociated with any $\gamma$-ray source we do not expect to detect LAT pulsation from these.
We have not yet detected LAT pulsation from the remaining MSP, indicating that possibly the pulsar is not associated with the 
$\gamma$-ray source.  
\section{GMRT high resolution southern sky survey}
We are conducting the GHRSS survey\footnote{http://www.ncra.tifr.res.in/ncra/research/research-at-ncra-tifr/research-areas/pulsarSurveys/GHRSS} 
from mid 2013. Till now we covered $\sim$1800 square degree of sky (62\% of the total target sky). Survey description and 
initial discovery of 10 pulsars are reported in \cite{Bhattacharyya_etal2016}. Left panel of Figure \ref{fig1} showes the 
GHRSS target sky and the pulsars discovered in this survey. Till now we have discovered 13 
pulsars (listed in Table 1) from this survey including one MSP, one pulsar with LAT pulsation and 
two mildly recycled pulsars. So our 
present discovery rate is 13 pulsars in 1800 deg$^2$, i.e. 0.007 pulsars per deg$^2$, which is one of the highest among 
off-galactic plane surveys. Following we provide description of some of the pulsars discovered with the GHRSS survey.\\
$\bullet$ We have discovered $\gamma$-ray pulsations aided by the radio timing from the GMRT, of our discovery PSR J0514$-$4408. This
pulsar has a double peaked profile at 322 MHz which evolves to a single peaked $\gamma$-ray profile. The relative alignment
of the $\gamma$-ray and radio peaks is unusual for a young, non-recycled pulsar.\\ 
$\bullet$ PSR J2144$-$5237 is a MSP with a multi-component profile having significant evolution with frequency.\\
$\bullet$ PSR J1516$-$43 is a mildly recycled pulsar in a 223 days orbit with a companion mass of 0.4 M$_{\odot}$. \\
$\bullet$ PSR J1255$-$46 is another mildly recycled pulsar.\\
We demonstrated rapid convergence in pulsar timing with a more precise position obtained with imaging and achieved 
long-term timing solutions for GHRSS  pulsars.
In addition to periodicity searches, the single pulse searches of the GHRSS survey can reveal sources like
Rotating Radio Transients (RRATs) and FRBs. The full survey is expected to
detect $\sim$ 4 fast radio bursts at fluence of 3 Jy-ms assuming a flat spectra. Even non-detection
of FRBs in 50\% of the survey yields a 2-sigma upper limit of 2000 events sky$^{-1}$ day$^{-1}$
at 322 MHz, which can put a constraining upper-limit on FRB spectral index.
Aided by the upgraded GMRT with increased sensitivity, we plan to complete the remaining part of the GHRSS survey with the
wider bandwidth (\cite{Roy17}).
\begin{table}
  \begin{center}
  \caption{Parameters of the pulsars discovered with the GMRT between 2012$-$2017 with targeted and blind search.}
  \label{tab1}
 {\scriptsize
  \begin{tabular}{|l|c|c|c|c||c|c|c|c|c|c|c|}\hline 
Pulsar name                & Period  & Dispersion measure & Type of pulsar/Orbital period         & Flux density$^\ddagger$ \\
                           & (ms)    & (pc~cm$^{-3}$)     & (day)                                 & (mJy)  \\\hline
PSR J0248$+$42             & 2.60    & 48.2               & isolated MSP                          & 1.9 \\
PSR J0418$-$4154$^\dagger$ & 757.11  & 24.5               & normal PSR             & 10.3                        \\
PSR J0514$-$4408$^\dagger$ & 302.2   & 15.4               & normal PSR             & 9.7  \\
PSR J0702$-$4956$^\dagger$ & 666.66  & 98.7               & normal PSR             & 15.7     \\
PSR J0919$-$42$^\dagger$   & 812.6   & 57.8               & normal PSR             & 6.4     \\
PSR J1120$-$3618           & 5.55    & 45.1               & MSP                    & 0.3 \\
PSR J1207$-$5050           & 4.84    & 50.6               & isolated MSP           & 0.5 \\
PSR J1227-4853             &1.686    & 43.4               & 0.28                  & 6.6 \\
PSR J1239$-$48$^\dagger$    &653.89  & 107.6              & normal PSR             & 0.4 \\
PSR J1255$-$46$^\dagger$    & 52.0   & 42.9               & mildly recycled        & 0.8 \\
PSR J1456$-$48$^\dagger$    & 536.81 & 133.0              & 15                     & 1.2   \\
PSR J1516$-$43              & 36.03  & 70.2               & mildly recycled        & 0.7 \\
PSR J1536$-$4948            & 3.08   & 38.0               & 62.5                   & 12 \\
PSR J1544$+$4937            & 2.16   & 23.2               & 0.12                   & 2.6 \\
PSR J1559$-$44$^\dagger$    & 1169.89 & 122.0             & normal PSR             & 1.7    \\
PSR J1646$-$2142            & 5.85    & 29.7              & isolated MSP                     & 0.7\\
PSR J1708$-$52$^\dagger$    & 449.62  & 102.6              & normal PSR                      & 1.4  \\
PSR J1726$-$52$^\dagger$    & 631.84  & 119.7              & normal PSR             & 0.7 \\
PSR J1828$+$0625$^\dagger$  &3.63     & 22.4              & MSP                    & 1.0 \\
PSR J1947$-$43$^\dagger$    & 180.94  & 29.9               & normal PSR                     & 4.7   \\
PSR J2144$-$5237$^\dagger$  & 5.04    & 19.0               & 10.58                    & 1.6 \\\hline
  \end{tabular}
  }
 \end{center}
 \scriptsize{
 {\it Notes:}
$^\dagger$ : GHRSS discoveries, $^\ddagger$ : Flux density is without primary beam correction. }
\end{table}

\end{document}